\documentstyle[epsf]{article}
\begin{document}

\title{A Solution to the Graceful Exit Problem in \\ Pre-Big Bang Cosmology}
\author{G F\ R\ Ellis\thanks{
email: ellis@maths.uct.ac.za}, D C\ Roberts\thanks{
email: roberts@maths.ox.ac.uk}, D Solomons\thanks{
email: deon@maths.uct.ac.za}, and P\ K\ S\ Dunsby\thanks{
email: peter@vishnu.mth.uct.ac.za} \\
Mathematics Department, University of Cape Town, \\
Rondebosch, Cape Town 7701, South Africa\thanks{
Present address of D Roberts: Department of Physics, University
of Oxford, Oxford, UK.}}
\date{3 March 2000}
\maketitle

\begin{abstract}
We examine the string cosmology equations with a dilaton potential in
the context of the Pre-Big Bang Scenario with the desired scale factor 
duality, and give a generic algorithm for obtaining solutions with 
appropriate evolutionary properties. This enables us to find pre-big bang 
type solutions with suitable dilaton behaviour that are regular at $t=0$, 
thereby solving the graceful exit problem. However to avoid fine
tuning of initial data, an `exotic' equation of state is needed that 
relates the fluid properties to the dilaton field. We discuss why such 
an equation of state should be required for reliable dilaton behaviour 
at late times.
\end{abstract}

\section{Introduction}

In this paper, we investigate the equations of string cosmology \cite
{Gasperini}, \cite{Lidsey} in the string frame, allowing for a dilaton
potential $V(\phi)$. The Pre-Big Bang Scenario is motivated by the search 
for
cosmological solutions with an $a(t)\rightarrow 1/a(-t)$ symmetry in the
scale factor $a(t),$ which implements an analogue of the T-duality symmetry
of M-theory. However one must distinguish between symmetries of the
equations and those of their solutions. We look at cases in which the {\it 
equations } have such a scale factor symmetry, when solutions may or may not
exhibit the same symmetry, and at cases in which the {\it solutions} obey
the scale factor symmetry, even if the equations do not. In the latter case
we obtain some solutions that seem to have most of the properties desired in
the Pre-Big Bang scenario, in that they have the desired scale factor
symmetry, the desired evolution of the dilaton field, and continuity at $
t=0$ of $a(t),\phi (t),\dot{\phi}(t)$, and the Hubble parameter $H(t)\equiv 
\dot{a}(t)/a(t)\ $ (but allowing a discontinuity in $\dot{H}(t)$ and 
$\ddot{\phi}(t)$ there, implying a corresponding discontinuity 
in $\partial V/\partial \phi $), thus providing a solution to the
graceful exit problem \cite{nogo,nogo1}. 
However, to obtain the desired dilaton behaviour at recent times, we
need to employ an `exotic' equation of state as discussed below. 

There are `no-go theorems' that exclude such regular 
transitions in the presence of a perfect fluid and Kalb-Ramond 
sources. A `lowest order' Einstein frame
analysis by \cite{nogo1} discusses graceful exit in 
generalized phase-space, and derives a set of necessary conditions for 
transition from a classical dilaton-driven inflationary pre-big bang
phase to a radiation-dominated era, joined at $t=0$ in a Planck epoch of 
maximal finite curvature $\dot{H}(t)$. They
show that a successful exit requires
violation of the null energy condition (NEC). Classical 
sources tend to obey NEC, but various new kinds of effective sources 
generating non-singular evolution have been considered that do not. Thus
failing invocation of higher order curvature terms,
some kind of exotic behaviour of matter is necessary in order
to obtain a graceful exit from the pre-big bang phase.

In this paper we follow Gasperini et al \cite{Gasperini} by working
in the string frame. The relation to the Einstein frame is left for another
paper. It should be made clear from the start that our solutions are
rather special in the spectrum of pre-big bang models; those we concentrate
on in the main show an exact scale factor duality in the solutions, and thus
we do not consider here the more exciting possibility of a phase of early
kinetic-dilaton dominated inflation which leads to an early phase which
is not radiation dual but is genuinely stringy inflationary vacuum. 
Nevertheless the set of solutions investigated here help to
understand the spectrum of possibilities available within the broad
Pre-Big Bang set of ideas.

\section{String Cosmology Equations}

\label{streqns} One can determine the general equations of string cosmology
by extremizing the lowest order effective action of dilaton gravity, 
\begin{equation}
S=-\frac 1{2\lambda _s^{d-1}}\int d^{d+1}x\sqrt{|g|}e^{-\phi }\left[
R+(\nabla \phi )^2-\frac 1{12}H^2+V(\phi )\right] \linebreak[4]+\int
d^{d+1}x \sqrt{|g|}L_m,
\end{equation}
where $\phi $ is the scalar dilaton, $H=dB$ (antisymmetric tensor field
strength), $V(\phi )$ is the dilaton potential, $\lambda _s$ is the
fundamental string length scale, and $L_m$ is the Langrangian density of
other matter sources. To derive string cosmology equations for the $d=3$,
homogeneous, isotropic, conformally flat background we will follow Gasperini 
\cite{Gasperini} in assuming $B=0$, a perfect fluid minimally coupled to the
dilaton, and a Bianchi I type metric (see Appendix C of \cite{Gasperini} for
details). Unlike Gasperini we assume $V(\phi )\ne 0,$ to obtain the string
cosmology equations in the following canonical form: 
\begin{equation}
H^2=\frac{e^\phi }6\rho +H\dot{\phi}+\frac V6-\frac{\dot{\phi}^2}6,
\label{friedman}
\end{equation}
\begin{equation}
\dot{H}+H^2=e^\phi (\frac p2-\frac \rho 3)-H\dot{\phi}+\frac{\dot{\phi}^2}3- 
\frac{V^{\prime }}2-\frac V3,  \label{ray}
\end{equation}
\begin{equation}
\ddot{\phi}=-3H\dot{\phi}+\dot{\phi}^2-V-V^{\prime }+e^\phi (\frac{3p}
2-\frac \rho 2),  \label{dilaton}
\end{equation}
where $V^{\prime }=\frac{\partial V}{\partial \phi }$. When combined, these
imply the standard energy conservation equation: 
\begin{equation}
\dot{\rho}=-3H(\rho +p)\;.  \label{energy}
\end{equation}
In a relationship analogous to that between the classical Friedmann equation
and Raychaudhuri equation, 
\begin{equation}
Eq.(\ref{friedman})~is ~the~ first~ integral~ of~ eq.(\ref{ray})~ provided
~that ~eq.(\ref{dilaton})~ and~ eq.(\ref{energy})~ hold.  \label{cons}
\end{equation}
These four equations will be the basis for the analysis in this paper.

One of the primary motivations for the pre-big bang scenario \cite{Veneziano}
is that when $V(\phi )=0$, these equations are invariant under the following
transformation: 
\begin{equation}
a(t)\rightarrow \hat{a}(t)=a^{-1}(t)  \label{a}
\end{equation}
provided that the dilaton transforms as 
$\phi \rightarrow \hat{\phi}=\phi -6\ln a$ 
and the energy density and pressure as $\rho \rightarrow \rho ^{\prime
}=a^6\rho ,\;p\rightarrow p^{\prime }=-pa^{-6}$. Thus if $a(t)$ is a
solution, so is $\hat{a}(t)$ for suitable $\phi ,\rho ,p.$ Since the string
cosmology equations are also invariant under time reversal symmetry,

\begin{equation}
a(t)\rightarrow \overline{a}(t)=a(-t)  \label{time_rev}
\end{equation}
the deceleration associated with standard post-big bang cosmology can be
associated with an accelerated evolution prior to the big bang by the
generalized transformation 
\begin{equation}
a(t)\rightarrow \tilde{a}(t)=a^{-1}(-t).  \label{dual}
\end{equation}
where $\tilde{a}(t)$ is a solution for suitable $\phi ,\rho ,p$ because $
a(t) $ is. The solution has T-duality symmetry if for each $t,$

\begin{equation}
a(t)=\tilde{a}(t)=a^{-1}(-t).  \label{duality}
\end{equation}

However, if we assume $V(\phi )\ne 0$ as in eqs.(\ref{friedman}-\ref{dilaton}
), then in general the equations are not invariant under the symmetry eq.(
\ref{duality}) even if the solutions are. We will look at both cases in what
follows, but generically allowing a potential that does not preserve the
symmetry. Note that if we assume matter with {\it the same } equation of
state before and after $t=0,$ then the matter equations also will not be
invariant under the scale factor symmetry. One has to decide what is more
physically meaningful: matter with a universal equation of state applicable
at all times, or that has a discontinuous equation of state that preserves
this symmetry. In what follows, we adopt the first option. We return to
discuss this choice in the conclusion.

\subsection{Flat Dilaton Potential with Exotic Equation of State}

To obtain equations of motion preserving the scale factor symmetry eq.(\ref
{duality}), we assume the simplest potential, namely a flat potential: 
\begin{equation}
V(\phi )=\kappa  \label{flat}
\end{equation}
where $\kappa $ is a constant, and then investigate the behaviour of the
universe. In order to reliably obtain proper limiting behaviour of the
dilaton, we assume that the equation of state 
\begin{equation}
p=\frac \rho 3+\frac 23e^{-\phi }\kappa  \label{eqnofstate}
\end{equation}
holds at all times (this choice, which is not invariant under the duality
symmetry, is discussed further in the following sections). One can
immediately see that at late times if $\phi \rightarrow $ constant, as we
will show follows from this choice, then this equation of state simply
reduces to radiation plus a constant.

We are interested in getting satisfactory dynamics for $H(t)$ and $\phi (t)$
, or equivalently for $\chi (t)\equiv \dot{\phi}$. To see when this occurs,
we manipulate the string cosmology equations (\ref{friedman}-\ref{dilaton})
subject to eqs.(\ref{flat},\ref{eqnofstate}) to obtain the two-dimensional
phase space with coordinates ($\chi $,$H$) governed by the following
equations: 
\begin{equation}
\dot{H}=\frac{\chi ^2}6-2H^2+\frac \kappa 6,
\label{dotH}
\end{equation}
\begin{equation}
\dot{\chi}=\chi (\chi -3H),
\label{dotchi}
\end{equation}

\begin{figure}
\epsffile{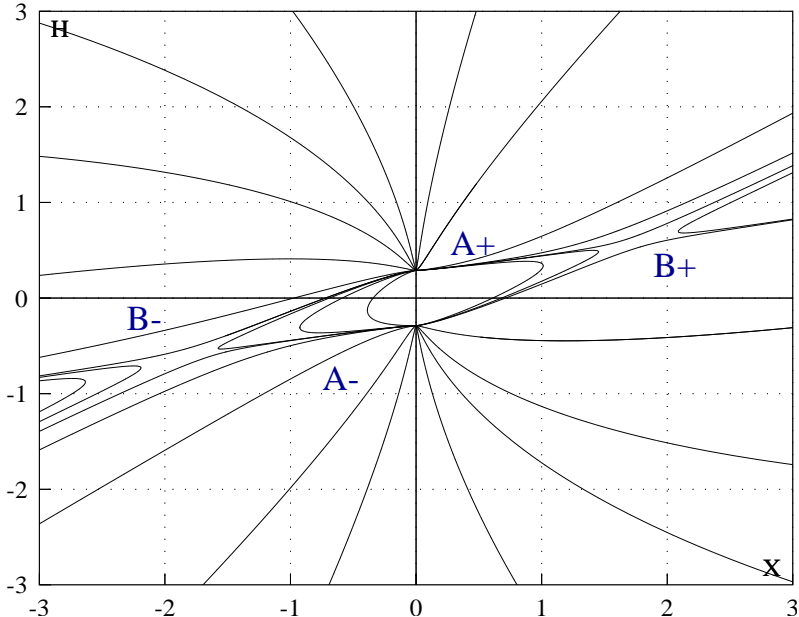}
\caption{Phase portrait representing the solution space of equations
  (\ref{dotH}-\ref{dotchi}) with $\kappa>0$.}
\end{figure}
the latter following because of choice (\ref{eqnofstate}). Having chosen the
constant $\kappa $, we can set initial conditions $(\chi _0,H_0)$ at $t=0$,
and then extend the solution to positive and negative values of $t$ by use
of these equations. For $\kappa <0$, there are no fixed points in the phase
plane, and on every trajectory both $H$ and $\chi $ diverge as $
|t|\rightarrow \infty $. For $\kappa =0$, i.e. no dilaton potential, there
is one fixed point at the origin, but for any initial condition (set at $t=0$
), $\chi $ and $H$ will diverge either as you run time forwards or run time
backwards.

The interesting dynamics is obtained when $\kappa >0$. There are then fixed
points at $A_{+}$: $(0,\sqrt{\kappa /12})$ (a source), $A_{-}$: $(0,\sqrt{
\kappa /12})$ (a sink), $B_{+}$: $(\sqrt{3\kappa },\sqrt{\kappa /3})$ (a
saddle point), and $B_{-}$:$\;(-\sqrt{3\kappa },-\sqrt{\kappa /3})$ (a
saddle point). In the phase plane depicted in Figure 1 above, we claim 
the initial conditions in the region I bounded by $A_{+}$, 
$B_{+}$ and $A_{-}$ and the separatrixes joining them give satisfactory
dynamics of both $H$ and $\chi $ which include $\chi \rightarrow 0$ as $
|t|\rightarrow \infty $, $\;\chi >0$ for all times so $\phi (t)$ is
monotonic, $H$ remains finite, and a ``bounce'' occurs that avoids the
initial singularity. Since region I is bounded by fixed points that have
coordinates proportional to $\sqrt{\kappa }$, increasing $\kappa $ will give
one a larger region of initial conditions that lead to a nonsingular
universe with proper dilaton dynamics. We can obtain a solution on the
boundary of region I (evolving along the line joining $A_{-}$ to $A_{+},$
which does not lie in I) that is invariant under symmetry (\ref{time_rev})
by setting $\chi _0=0,\;H_0=0$ at $t=0,$ but this solution, given explicitly
by $a(t)=\cosh ^{1/2}(\sqrt{\kappa /3}t),$ is not invariant under the
symmetry (\ref{duality}). A drawback of all these models is that inflation
will not stop at $t>0$, but as discussed in the conclusion, the string
cosmology equations derived in section \ref{streqns} do not apply to the
present cosmological regime without modification, so it is possible that a
radiation dominated evolution started after the time when these equations no
longer apply. In any case this gives a specific family of solutions where
the equations display the desired symmetry (\ref{duality}) but the solutions
do not - which is not very surprising, given the prevalence of broken
symmetries in physics.

\section{Obtaining Desired Dynamics From a Dilaton Potential}

In this section, we generalize the method introduced by Ellis and Madsen 
\cite{Ellis} through which they obtain a classical scalar potential
associated with a specified $a(t)$ in the standard gravitational equations.
No field has been observed that coincides with a dilaton potential $V(\phi )$
, so we assume that it is a freely disposable function. We show that by
suitable choice of $V(\phi )$ one can obtain almost any behaviour for $a(t),$
or alternatively for $\phi (t)$. We first present an algorithm for
determining $V(\phi )$ from a desired $a(t)$ or a desired $\phi (t)$, and
then present an analytically smooth solvable example. This solution
illustrates our main point, but has little physical relevance (although it
does satisfy the symmetry (\ref{duality})). In the following section we use
these methods to obtain two solutions that resemble the standard ``pre-big
bang scenario'', but with continuity of $a(t)$ and $H(t)$ and with
satisfactory dynamics of $\phi (t)$. The associated dilaton potentials are 
{\it ad hoc} because they are derived from the desired behaviour of the
universe, rather than from a field theory model; as discussed 
in many inflationary and quintessence studies, see e.g. \cite{Lidsey1},\cite
{Saini}.

\subsection{The Algorithms}

\label{algorithm}

We proceed by providing the following {\it general algorithm }for
determining a dilaton potential $V(\phi )$ that produces a desired $a(t)$:

{\bf 1)} Specify a desired monotonic function for the scale parameter $a(t)$
, consequently determining $H(t)$ and $\dot{H}(t)$,

{\bf 2)} Choose an equation of state and solve for $\rho (a)$ from eq.(\ref
{energy})\footnote{
If the equation of state is a function of $V$ or $V^{\prime}$, then you will
have to eliminate these quantities using eqs.(\ref{friedman}) and (\ref{ray}
) before solving eq.(\ref{energy}).}; as $a(t)$ is known, this determines $
\rho (t)$.

{\bf 3)} Eliminate $V$ and $V^{\prime }$ from eq.(\ref{dilaton}) by use of
eqns.(\ref{friedman}) and (\ref{ray}) to obtain a differential equation
relating $H(t)$, $\phi(t)$, $\rho(t)$, and their time derivatives. 

{\bf 4)} Solve the equation obtained in step 3) for $\phi (t).$

{\bf 5)} Substitute the now known functions $\phi (t)$, $\rho (t)$, and $
H(t) $ and $a(t)$ into the rearranged version of eq.(\ref{friedman}) 
\begin{equation}
V(t)=6H^2-e^\phi \rho -6H\dot{\phi}+\dot{\phi}^2  \label{V}
\end{equation}
to obtain $V(t)$.

{\bf 6)} Invert $\phi (t)$ to obtain $t(\phi )$, and

{\bf 7)} Transform $V(t)$ as follows: $V(t)=V(t(\phi ))\Rightarrow V(\phi )$
. This is possible for each range of $t$ on which $\phi (t)$ is monotonic
(if it is not monotonic on some range of $t$, in general $V(\phi )$ will not
be well-defined because it will not be single valued for the corresponding
values of $\phi $).

Thus, provided $\phi (t)$ determined from step 3) 
is monotonic, we find a $V(\phi )$ that corresponds to a given monotonic
function $a(t)$. Because we have now satisfied eqs.(\ref{energy}), (\ref
{friedman}) and 
the equation obtained in step 3), the latter depending essentially on eq.(
\ref{dilaton}), it follows from statement (\ref{cons}) that eq.(\ref{ray})
will be true also, so we have satisfied all the equations of the theory for
this matter description (c.f. \cite{Ellis}); hence we have a solution of the
desired form.

Alternatively, we can give an algorithm for determining a dilaton potential $
V(\phi )$ that produces a desired dilaton evolution $\phi (t)$\footnote{
It is important to note that one has freedom to choose only $a(t)$ or $
\phi(t)$, not both.} by proceeding in the same way as above, except for
minor changes: replace step {\bf 1)} by

{\bf 1')} specify the desired monotonic function for the dilaton, $\phi (t),$
\\in step {\bf 2)}, leave $\rho $ in the form $\rho (a),$ and replace step 
{\bf 4)} by

{\bf 4')} solve the equation obtained in step 3) for $a(t)$ (or for $H(t)$).
\\The rest of the algorithm is as before.

Finally, note that we can carry out these procedures piecewise: for example
we can specify $a(t)$ for some range of $t$ and $\phi(t)$ for some adjoining
range of $t$, or different behaviours for $a(t)$ for different ranges of $t$
, then join the solutions together, ensuring that $a(t)$, $H(t)$, $
\phi(t) $ and $\chi(t)$ are continuous where these ranges meet.

\subsection{Exponential Scale Factor Behaviour with No Matter}

To demonstrate the procedure, we give a simple analytically solvable example
with a pure scalar field, i.e. $\rho =p=0$. Consider an exponential
expansion as in classical inflation, 
\begin{equation}
a(t)=e^{wt}\Rightarrow H=w, ~\dot{H}=0,
\end{equation}
where $w$ is a positive constant. This solution has the desired symmetry (
\ref{duality}).

In this case the differential equation 
for $\phi (t)$ takes the form: 
\begin{equation}
\ddot{\phi}=H\dot{\phi}  \label{phi1}
\end{equation}
Using eq.(\ref{phi1}) and eq.(\ref{V}) we obtain 
\begin{equation}
\phi (t)=\phi _0+\frac{\dot{\phi}_0}w(e^{wt}-1),  \label{phi(t)}
\end{equation}
a monotonic function as required, and 
\begin{equation}
V(t)=6w^2-6\dot{\phi}_0we^{wt}+\dot{\phi}_0^2e^{2wt}.  \label{V(t)}
\end{equation}
After inserting the inverted eq.(\ref{phi(t)}), 
\begin{equation}
t(\phi )=\frac 1w\log \left[ \frac w{\dot{\phi}_0}\left( \phi -\phi _0+\frac{
\dot{\phi}_0}w\right) \right] ,
\end{equation}
into eq.(\ref{V(t)}), one obtains 
\begin{equation}
V(\phi )=w^2(\phi -3-\phi _0+\frac{\dot{\phi}_0}w)^2-3w^2
\end{equation}
which is simply a quadratic potential plus a constant. Clearly the behaviour
for $\phi (t)$ is unphysical since $\phi (t)\rightarrow \infty $ instead of
asymptoting to a constant. However, this gives a transparent example where
even though the scale factor symmetry (\ref{duality}) is broken in the
equations (because $V(\phi )$ is not constant), the solution obeys that
symmetry.

\section{``Pre-big Bang'' Behaviour}

In this section we try to use the methods just explained to obtain solutions
that resemble the ``pre-big bang scenario'' but with satisfactory dynamics
of $\phi (t)$ and a continuous transition from the pre-big bang to post-big
bang phases. In these examples, we seek solutions that evolve from a string
perturbative vacuum, i.e. $H\rightarrow 0$ and $e^\phi \rightarrow 0$ (no
interactions), to the present scenario where $e^\phi $, which acts as the
coupling constant, asymptotes to a constant. We will assume the following
behaviour of the universe: 
\begin{equation}
a(t)=(t+1)^{\frac 12},\;t\geq 0\Rightarrow H(t)=\frac 1{2(t+1)}  \label{t}
\end{equation}
determining $a(t)$ for $t\geq 0,$ and by the symmetry (\ref{duality}) 
\begin{equation}
a(t)=(-t+1)^{-\frac 12},\;t\leq 0\Rightarrow H(t)=\frac 1{2(-t+1)}
\label{t-}
\end{equation}
determining $a(t)$ for $t\leq 0$. Both $a(t)$ and $H(t)$ are continuous at $
t=0$ with $a(0)=1,\;H(0)=1/2$, but $\dot{H}(t)$ is not continuous there.

This behaviour, which is essentially radiation dominated evolution of the
universe for positive times and power-law inflation for negative times, is
motivated by the ``pre-big bang'' scenario introduced in \cite{Veneziano},
and exactly obeys the scale factor symmetry (\ref{duality}). Note that we
have shifted the origin of time in each branch from that customarily used,
in order to get a smooth evolution through $t=0$; this of course makes no
difference to the desired physical behaviour, for we can choose the origin
of time to be wherever we want (and the equations are invariant under time
translation $t\rightarrow t^{\prime }=t+c).$ Although the power law
inflation ends with the scale-factor value $a(0)=1,$ required by continuity
together with the symmetry (\ref{duality}), the solution has sufficient
inflation for any purpose because it involves an infinite number of
e-foldings (it starts with the asymptotical value $a=0$ as $t\rightarrow
-\infty ).$

\subsection{Pre-Big Bang behaviour with radiation equation of state}

First we assume the radiation equation of state holds at all times, that is, 
\begin{equation}
p=\frac \rho 3,  \label{rad}
\end{equation}
which, using eqs.(\ref{energy}) and (\ref{t},\ref{t-}), implies 
\begin{equation}
\rho (\pm t)=\rho _0(\pm t+1)^{\mp 2}  \label{rho}
\end{equation}
where $\rho _0$ is a positive constant and `$+t$' refers to the post-big
bang era, `$-t$' to the pre-big bang era. Notice that both $\rho $ and $\dot{
\rho}$ are continuous at $t=0$.

The equation for $\phi $ now takes the form 
\begin{equation}
\ddot{\phi}=\frac 23e^\phi \rho +H\dot{\phi}+2\dot{H}.  \label{phi2}
\end{equation}
Substituting in eqs. (\ref{t}) and (\ref{rho}), we could not find an
analytical solution to eq.(\ref{phi2}), so we investigate the three
dimensional phase space with coordinates $(t,\phi ,\chi ),$ given from
eqs. (\ref{phi2},\ref{t},\ref{t-},\ref{rho}) by 
\begin{equation}
\dot{\phi}=\chi ,~~~\dot{\chi}=\frac 23e^\phi \rho _0(\pm t+1)^{\mp 2}+\frac
\chi {2(\pm t+1)}\mp \frac 1{(\pm t+1)^2},  \label{ev}
\end{equation}
where the top sign holds for $t>0$ and the bottom sign for $t<0.$ We can set
initial data at $t=0$, and then investigate the phase plane orbits as we run
the trajectory forward and backwards in time in such a way that $\chi $ and $
\phi $ are continuous through $t=0$. Then $\dot{\chi}$ is discontinuous
there, but we have no problem in joining the solutions for $t>0$ and $t<0$.

For $t>0$, there is an exceptional integral curve $\gamma (t)$ given by $(t, 
\tilde{\phi}_0$,$0$), where $\tilde{\phi}_0\equiv $ ln$(\frac 3{2\rho _0})$;
this is the only integral curve with a fixed value of $\phi $ and $\chi $.
Note that setting $\phi _0$ and $\chi _0$ determines the initial point in
the phase space, and specifying $\rho _0$ determines the location of this
exceptional curve. In the 2-dimensional sub-spaces $t=const$ with
coordinates ($\phi ,\chi ),$ the curve $\gamma (t)$ has coordinates $(\tilde{
\phi}_0,0)$ for all $t,$ and represents a set of saddle points parametrised
by $t$. To get exactly the desired dilaton dynamics in the future ($\chi >0$
, $e^\phi \rightarrow $ constant $\Rightarrow \chi \rightarrow 0$ as $
t\rightarrow \infty $), one must restrict the initial conditions ($\phi _0$,$
\chi _0$) to start precisely on the stable branch of these saddle points,
which intersects the surface $t=0$ in a single curve $(0,\gamma _{+}(\chi
),\chi )$ passing through the exceptional point $\gamma _0=(0,\tilde{\phi}
_0,0)$ (for more details, see Appendix A). However there is actually
slightly more freedom than this in finding physically relevant initial
conditions because if a trajectory starts close enough to the stable branch
(but not exactly on it), then the trajectory will stay close to the fixed
point for an arbitrarily long period of time before $\phi $ and $\chi $
diverge, and this may suffice for physical purposes even if the solution
eventually diverges (cf. the discussion of intermediate isotropisation in 
\cite{wain}). Nevertheless, the physically relevant set of solutions is very
unstable and requires very precise fine tuning, in order to obtain the 
desired dilaton dynamics, lying in a small open neighbourhood 
${\cal D_{+}}$ of the curve $\phi _0=\gamma _{+}(\chi _0)$ in the
initial data set at $t=0$. Indeed we have found it very difficult to 
obtain numerical solutions with the desired behaviour because of this 
instability.

For $t<0$ there are no points with a fixed value of $\phi$ and $\chi$
(because we assume $\rho _0>0)$. To get the desired dilaton dynamics in the
past ($\chi >0$, $e^\phi \rightarrow 0$ as $t\rightarrow -\infty $) one must
further restrict the initial conditions, the problem being that eq.(\ref{ev}
) is an inhomogeneous equation for $\chi $ with a time-varying source
function (albeit a source function that decays away as $t\rightarrow \pm
\infty )$. We can obtain the desired behaviour if $\;y_0=\frac
23e^{\phi_0}\rho _0\ll 1,$ i.e. $\phi_0\ll \ln (\frac 3{2\rho _0})$ (details
are given in Appendix A). This is a sufficient condition; there will be a
wider domain ${\cal D_{-}}$ of initial data at $t=0$, containing this set,
that will ensure that at early enough times the desired behaviour is
attained.

To get a satisfactory solution for all time, for a given choice of $\rho _0,$
one must set the initial conditions to lie in both ${\cal D}_{{\cal +}}$ and 
${\cal D_{-}}$, so the crucial issue is whether they intersect or not. We
have not attained finality on this point. It may be that the `no-go'
theorems with a potential \cite{nogo} imply they do not intersect, but this
implication is not entirely clear, as the conditions of those theorems may
not correspond precisely to the conditions we contemplate here. If they do
intersect, we can attain the desired behaviour $\chi \rightarrow 0$ and $
e^\phi \rightarrow 0$ when time runs backwards as well as $e^\phi
\rightarrow const$ as time runs forward and in principle, one can obtain a
continuous $V(\phi )$ associated with the unstable solution described above
because every function is continuous on the right hand side of eq.(\ref{V}).
Furthermore, $\phi (t)$ is monotonic and continuous, and therefore
invertible, so one can complete Step 6 of the algorithm set out in section 
\ref{algorithm}. However attaining such solutions will require extreme
fine-tuning of the initial data, and this is very difficult to do because
one does not know where the stable branch of the saddle point
intersects $t=0$. Thus if such solutions do exist, the extreme fine 
tuning required for their initial data make them seem impracticable 
as cosmologies despite their other desirable properties.

\subsection{``Pre-big Bang'' Behaviour with Exotic Equation of State}

Finally, we assume the identical ``pre-big bang'' behaviour of the last
example (eqs.(\ref{t},\ref{t-})), but we obtain a stable solution with a
different equation of state. The instability in the last example arises
because of our choice of the equation of state, as can be seen by inspection
of eq.(\ref{dilaton}), which we write now as 
\begin{equation}
\dot{\chi}=-3H\chi +\chi ^2+\beta ,  \label{dilaton2}
\end{equation}
where 
\begin{equation}
\beta \equiv -V-V^{\prime }+e^\phi (\frac{3p}2-\frac \rho 2).  \label{beta}
\end{equation}

As mentioned before, we want to obtain $e^\phi \rightarrow $ constant, i.e. $
\chi \rightarrow 0$, at late times, which implies $\beta \rightarrow 0$ in
eq.(\ref{dilaton2}). If we choose the radiation equation of state as in the
last example, (eq.(\ref{rad})), then $\beta =-V-V^{\prime }$. Therefore,
requiring $\beta \rightarrow 0$ as $t\rightarrow \infty $ puts a heavy
restriction on the dilaton potential, namely $V\rightarrow e^{-\phi }$ at
late times. Consequently, there is a fine-tuning problem if you use the
radiation equation of state.

In the present example, we assume $\beta =0$ for all times, which from eq.(
\ref{dilaton2}) demands the exotic equation of state, 
\begin{equation}
p=\frac \rho 3+\frac 23e^{-\phi }(V+V^{\prime }),  \label{exotic}
\end{equation}
at all times (note that eq.(\ref{eqnofstate}) is the special case resulting
when $V^{\prime }=0$). Using this equation of state implies, 
\begin{equation}
\rho (t)=\int 2He^{-\phi }(12H\dot{\phi}+6\dot{H}-3\dot{\phi}^2)dt
\label{ccons}
\end{equation}
which allows the density to go through zero and become negative. We discuss
this equation further in the {\it Conclusion}.

The differential equation that relates $H(t)$ to $\phi (t)$ is simply eq.(
\ref{dilaton2}) with $\beta =0$, 
\begin{equation}
\ddot{\phi}=-3H\dot{\phi}+\dot{\phi}^2,
\end{equation}
For arbitrary $a(t),$ this can be solved (with $a_0^{\ }=1$ and $\chi
_0\equiv \dot{\varphi}_0$) by

\begin{equation}
\exp (\phi _0-\phi (t))=1-\chi _0\int_0^ta^{-3}(t)dt.  \label{soln}
\end{equation}
For the specific case given by eqs.(\ref{t},\ref{t-}) we obtain from this
the analytical solution 
\begin{equation}
\phi (t)=+\phi _0-\ln \left| 1-2\chi _0[1-(1+t)^{-{\frac 12}}]\;\right| 
\label{phi+}
\end{equation}
for $t>0$ and 
\begin{equation}
\phi (t)=+\phi _0-\ln \left| 1-{\frac{2\chi _0}5[1-}(1-t)^{{\frac 52}
}]\;\right|   \label{phi-}
\end{equation}
for $t<0$. Inverting eq.(\ref{phi+}) we obtain 
\begin{equation}
t(\phi )=\frac{4\chi _0^2e^{2(\phi -\phi _0)}}{[(1-2\chi _0)e^{\phi -\phi
_0}-1]^2}-1  \label{solnst}
\end{equation}
and inverting eq.(\ref{phi-}) we obtain 
\begin{equation}
t(\phi )=1-\left[ \frac{5e^{\phi -\phi _0}-5+2\chi _0}{2\chi _0}\right]
^{2/5}  \label{solnend}
\end{equation}
Now we can solve eq.(\ref{ccons}) to obtain $\rho (a)$ and so $\rho (t)$
(see Appendix B for one particular case), and substitute our results into
eq.(\ref{V}) to obtain the dilaton potential $V(\phi )$ that is associated
with our specified ``pre-big bang'' behaviour. This is straightforward but
tedious, and results in very complex analytic expressions (the real
complexity coming through the expressions for $\rho (t)$ that occur
consequent on the choice of the exotic equation of state). Rather than
giving these analytic expressions, we give a graph of the potential for one
particular case below. \\ 

To discuss the relevant initial conditions, it is instructive to look at the
phase plane (Figure 2 above) with coordinates ($t,\chi$
), where $\chi =\dot{\phi}$ is governed by the equation: 
\begin{equation}
\dot{\chi}=-\frac 3{2(\pm t+1)}\chi +\chi ^2  \label{chi}
\end{equation}
where we again use $+$ to represent $t>0$ and $-$ to represent $t<0$. One
can easily see that $\chi =0$ $(\Rightarrow \dot{\chi}=0)$ is an attractor,
and represents a physically uninteresting solution with $\phi =const$. Also $
\chi =\frac 3{2(\pm t+1)}$is a nullcline, characterising the other points
where $\dot{\chi}=0$. This curve starts at $(0,\frac32)$ and drops
symmetrically away to zero as $t\rightarrow \pm \infty .$ Now we can solve
eq.(\ref{chi}) analytically for $t>0,$ finding

\begin{figure}
\epsffile{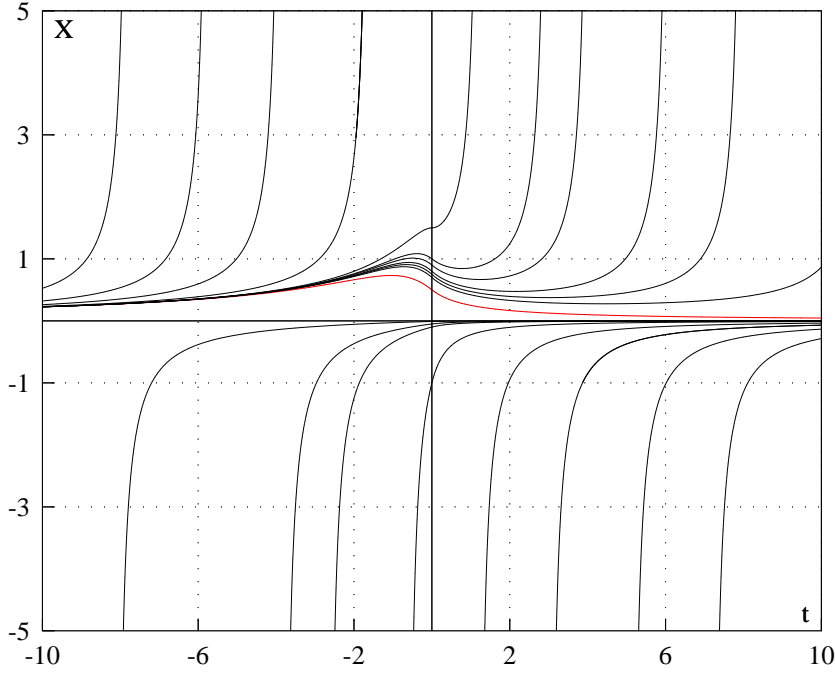}
\caption{Phase portrate representing the solution space of 
equation (\ref{chi}).}
\end{figure}

\begin{equation}
\chi =\frac 1{2(t+1)(1+C_{+}\sqrt{t+1})}  \label{chi1}
\end{equation}
where $C_{+}=(\frac 1{2\chi _0}-1)$ is positive iff $\;\chi _0<1/2.${\bf \ }
The separatrix between the solutions that diverge and those that go
asymptotically to zero as $t\rightarrow \infty $ is the special solution
with $C_{+}=0$ which goes through $(0,\frac12),$ that is,

\begin{equation}
\chi =\frac 1{2(t+1)}  \label{chi2}
\end{equation}
which itself goes to zero as $t\rightarrow \infty .$ If we specify the
initial conditions at $t=0$ such that $\phi _0$ is free and $0<\chi _0<\frac
12\Leftrightarrow C_{+}>0$, then as we run the trajectories forward in time $
\chi \rightarrow 0.$ In this case, for large positive values of $t$, eq.(\ref
{chi1}) will be approximately

\begin{equation}
\chi =\frac 1{2C_{+}\;t\sqrt{2t}\ }\ >0  \label{chi1+}
\end{equation}
(note that $\phi (t)$ is monotonic for $t>0$ because $\chi >0$ on these
trajectories). Let $T_{+}$ be such that eq.(\ref{chi1+}) is valid for all $
t>T_{+}>0$. Then for $t>T_{+},$ 
\begin{equation}
\phi (t)\simeq \int_{T_{+}}^t\frac 1{2C_{+}t\sqrt{2t}\ }dt+\phi
_{T_{+}}=\frac 1{C_{+}\sqrt{2}\ }[T_{+}^{-1/2}-t^{-1/2}]+\phi _{T_{+}}.
\end{equation}
Thus as $t\rightarrow \infty $, for all $\chi _{_0},$ $\phi (t)\rightarrow $
a constant value, say $\phi _\infty ,$ and $\exp \phi (t)\rightarrow \exp
(\phi _\infty) $. (Note that it is essential to check this result even
though $\chi \rightarrow 0$, cf. the discussion below of what happens as $
t\rightarrow -\infty ).$ If we specify the initial conditions at $t=0$ such
that $\phi _0$ is free and $\frac 12<\chi _0\Leftrightarrow C_{+}<0$, as we
run the trajectories forward in time then $\chi \rightarrow \infty $ as $
t\rightarrow t_0$ given by $1+C_{+}\sqrt{t_0+1}=0,$ that is $t_0=\frac{
(2\chi _0)^2-1}{(2\chi _0-1)^2}.$ In this case for large values of $\chi ,$
eq.(\ref{chi}) can be approximated as follows:

\begin{equation}
\chi \gg \frac 3{2(t+1)}\Rightarrow \dot{\chi}\simeq \chi ^2\Rightarrow \chi
\simeq 1/(t-t_0).  \label{chismall}
\end{equation}
The solution diverges as $t\rightarrow t_0$ and the approximation eq.(\ref
{chi1+}) never applies. This behaviour conforms to that implied by eq.(\ref
{phi+}), and may be seen clearly on the phase plane.

If we run the trajectories backward in time, starting from initial data with 
$\chi _0>0$, they will cross the nullcline and then drop to zero, never
becoming negative because $\chi =0$ is an exceptional solution of the
equations. Then $\phi (t)$ is monotonic for $t<0$ also because $\chi >0$ on
these trajectories. Solving eq.(\ref{chi}) analytically for $t<0$ gives

\begin{equation}
\chi =-\frac 52\frac{(t-1)\sqrt{-t+1}}{(t^2-2t+1)\sqrt{-t+1}+C_{-}}
\label{neg}
\end{equation}
where $C_{-}=(\frac 5{2\chi _{_0}}-1).$ This expression goes to zero for all 
$C_{-}>-1$, corresponding to $\chi _{_0}>\;0\ $ (note that it does not
matter if $C_{-}$ is positive or negative). For large negative $t$ its
value, for all $C_{-},$ will be approximately 
\begin{equation}
\chi =d\phi /dt\simeq -\frac 5{2t\ }.  \label{neg1}
\end{equation}
Let $T_{-}$ be such that eq.(\ref{neg1}) is valid for $t<T_{-}<0.$ Then for $
t<T_{-}$, 
\begin{equation}
\phi (t)\simeq \int_{T_{-}}^t(-\frac 5{2t\ })dt+\phi_{T_{-}} =\frac 52\ln ( 
\frac{T_{-}}t)+\phi_{T_{-}}-~~ \Rightarrow ~~\exp \phi (t)\propto (\frac{
T_{-}} t)^{5/2}
\end{equation}
Thus as $t\rightarrow -\infty $, for all \ $\chi _{_0},$ $\phi
(t)\rightarrow -\infty $ even though $\chi \rightarrow 0,$ and $\exp \phi
(t)\rightarrow 0,$ which is the dynamics we desire \cite{Gasperini}, and
indeed is indicated already by eq.(\ref{phi-}). The value of $C_{-}$
corresponding to the separatrix eq.(\ref{chi2}) is $C_{-}=4,$ which does not
give any special behaviour for $t<0.$

Typical results of the integrations for this case are given in Figures
3-5 below.

\begin{figure}
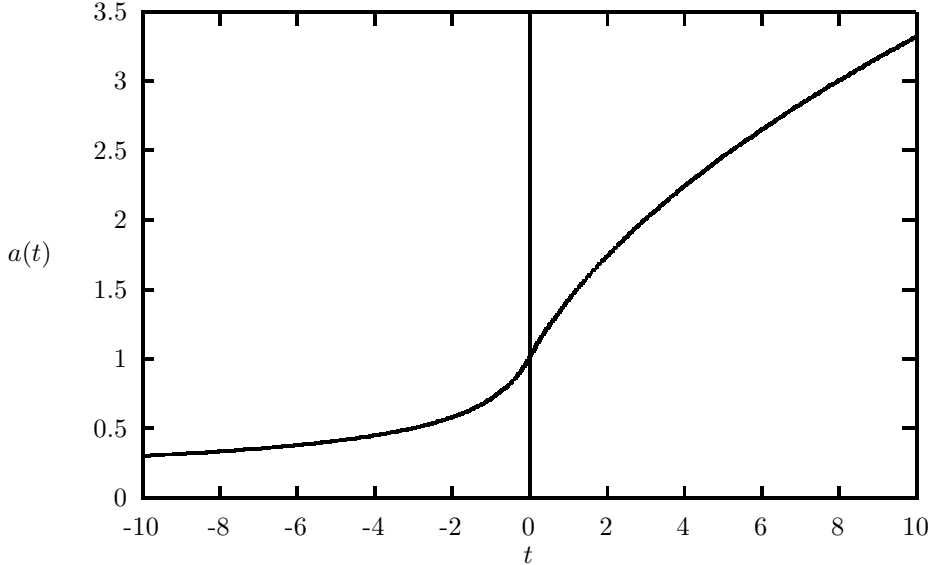

\input fig3.tex
\caption{The evolution of the scalefactor $a(t)$ as a function of
time $t$, with $a(0)=1$, over the time interval $[-10,10]$. For negative
times $t\leq 0$, there is power-law inflation, $t\geq 0$, followed by a
radiation dominated phase of expansion for positive time $t\geq 0$.}
\end{figure}

\begin{figure}
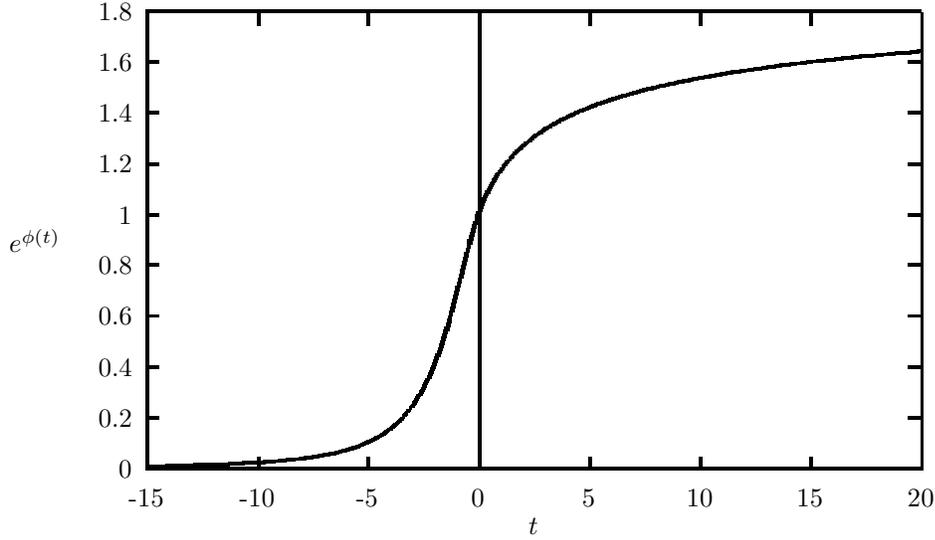

\input fig4.tex
\caption{The function $\exp{(\phi(t))}$ as a function 
of time $t$, with $a(0)=1$, $\phi(0)=0$ and $\chi(0)=0.25$. $\exp{(\phi(t))}$ 
increases monotonically from $0$ at time $t=-\infty$ to $2$ at $t=+\infty$.}
\end{figure}

In summary, one
gets a stable solution for $0<\chi _0<1/2$, as one can see from the phase
plane, with good ``pre-big bang'' behaviour and the desired dynamics for $
\phi (t){\bf \ }$ for both large and small $t.$ The shape of the potential
is a bit unusual, but results directly from the specific requested `pre-big
bang' evolution eqs.(\ref{t},\ref{t-}) and the chosen initial conditions.
Smoothing out that behaviour at $t=0,$ so that the solution departs from the
`radiation' form eq.(\ref{t}) at very early times while preserving the
symmetry (\ref{duality}), will result in a smoothed out potential $V(\phi )$
; we can choose $a(t)$ in this way so that $H(t)$ and hence $V(\phi )$ are
continuous at $t=0.$ Initial conditions can be set so that the matter has
the desired late time behaviour: $p/\rho \rightarrow 1/3,\rho \rightarrow 0;$
however it then has unusual behaviour at early times in that both $\rho $
and $h\equiv \rho +p$ go negative for some values of $t<0$. It is unclear if
this should be regarded as a serious defect of the model or not, remembering
that with the unusual equation of state adopted, the properties of matter
are different than usual, and in particular the speed of sound will no
longer be given by the usual expression. This needs further investigation.
What is clear is that 
these solutions are not physically reliable as $t\rightarrow +\infty $ (see
below), and they will have to be joined on to some other solution to give an
adequate model of the universe with ordinary matter behaviour at late times.
However, as discussed below, that problem occurs in the entire family of
pre-big bang models, and so is not restricted to the models considered here. 

\begin{figure}
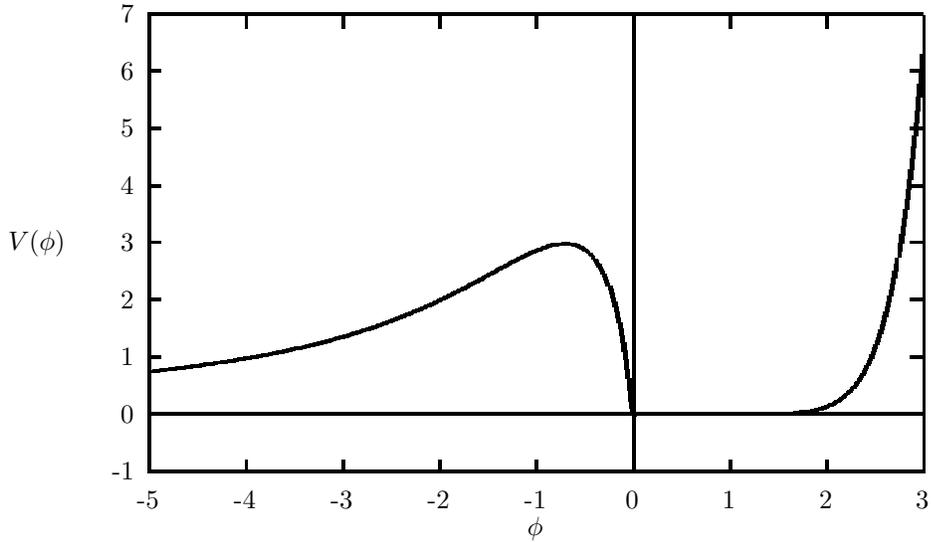

\input fig5.tex
\caption{The dilaton potential $V(\phi)$ as a function of time $
\phi$. We assume that $a(0)=1$, $\phi(0)=0$ and $\chi(0)=0.25$ and take the
density $\rho(t)$ to have value $\rho(\infty)=0$ at time $t=\infty$. The
potential $V(\phi)$ is continuous at all times, but non-differentiable at $
\phi=0$. For $\phi\rightarrow-\infty$, $V(\phi)$ is asymptotically zero. To
the right of $\phi=0$, the potential starts at $V\approx -0.005$ and goes to
zero from below as phi goes to $\ln{2}$, then increases to $+\infty$ as $
\phi\rightarrow\infty$. Around $\phi=\ln{2}$, both $V(\phi)$ and its
gradient $V^{\prime}(\phi)$ are %is flat and 
zero. As time $t\rightarrow 
+\infty$, the dilaton field asymptotes to a constant value of $\ln{2}$ 
in our model.
The dilaton potential $V(\phi)$  approximates  
a fixed value of $0$ as $\phi\rightarrow \ln{2}$ asymptotically 
for large positive times.}
\end{figure}

\section{Discussion}

We have given examples making very clear the distinction between the
equations and the solution having the desired `pre-big bang' symmetry. We
have given a broad method of attaining desired string cosmology solutions
when there is a dilaton potential $V$ not equal to zero, and used it to obtain
`Pre-Big Bang' solutions that seem to have close to the desired properties.
In the first case considered, choice of the exact radiation equation of
state (\ref{rad}) at all times leads to a very unstable situation where
extreme fine-tuning of initial conditions is required to attain the desired
results, and indeed there may be no initial data leading to the desired
behaviour in both the forward and backwards directions of time. In the
second case we impose an `exotic' equation of state (\ref{exotic}) that
links the fluid behaviour to the potential in a way that generalises the
perfect fluid equation of state, and we obtain solutions of the desired type
without the need for fine tuning the initial data set at $t=0.$

This equation of state looks strange, and the resulting matter behaviour is
certainly unusual, but we have no solid handle to use in restricting
equations of state in this early era; and we suggest that {\it it is
essential to choose such an equation if one wants the solution to reliably
tend to the `classical' form at late times}. This is because of the form of
the equation for $\ddot{\phi}$; if we do not set $\beta =0$, where $\beta $
is defined by eq.(\ref{beta}) then almost always that desired classical
state will not be attained, because of eq.(\ref{dilaton2}); but setting $
\beta =0$, which leads to the desired behaviour, leads immediately to our
`exotic' equation of state. Insofar as that equation of state and resulting
behaviour is unsatisfactory, this indicates that {\it there is a problem
with the form of the equation for} $\ddot{\phi}$, which comes directly from
the standard variational principle employed in the context of the pre-big
bang scenario. The remedy probably lies in finding other scenarios
with alternative forms of the variational principle, leading to other
equations for $\ddot{\phi}$.

This is also indicated because the present form of the equations does not
accommodate ordinary matter, the point being that the above analysis applies
even if there is no dilaton potential. Suppose $V=0;$ then eq.(\ref{dilaton2}
) remains true, but now 
\begin{equation}
\beta =e^\phi (\frac{3p}2-\frac \rho 2),  \label{beta1}
\end{equation}
so a reliable approach of the dilaton to a classical solution at late times,
requiring $\beta =0,$ demands the radiation equation of state
(\ref{rad}); a baryon dominated epoch is not allowed\footnote{
Although of course by the algorithm given above, we can simulate a matter
dominated phase by suitable choice of the potential $V.$}. This is usually
dealt with by stating that eqs.(\ref{friedman}-\ref{energy}) don't apply at
late times in the history of the universe - a different set of equations are
to be used then, and the solutions for early times obtained from eqs.(
\ref{friedman}-\ref{energy}) must be suitably joined on to that late time
evolution. However given the vision of M-theory as representing the
fundamental theory of gravity, it should be able to describe that epoch too;
this apparently requires some modified scenario and associated variational
principle (note that although we have discussed the issue in the string
frame, it also arises in essentially the same form in the Einstein frame).
In any case, whether one accepts this argument or not, given the standard
variational principle and equations, we argue that the `exotic' equation of
state implied by setting $\beta =0$ is {\it necessary} to give the desired
behaviour; when adopted, it enables obtaining that behaviour reliably (i.e
it eliminates the need for extreme fine-tuning of data set at $t=0$).

However one should note here that we have perhaps been somewhat extreme
in imposing this equation of state at all times. It is only really needed,
on our approach, near the time of the turnaround, and one could obtain far 
more general behaviours by modifying what we have here in that light; 
what is required
is that the quantity $\beta$ must go to zero in the period when the
dilaton is stabilised. It has also been pointed out
to us that it is not clear why the deviation from its vanishing point should 
be absorbed completely in the pressure, and then promoted into the 
conservation equation; other models of the transition 
\cite{trans1,trans2,trans3} 
successfully stabilise the dilaton at late
times without this requirement, with suggestions for classical and quantum
corrections in the effective action taking the place of the exotic fluid.
Hence our proposal must just be seen 
as one of a range of possibilities in this regard. 

Because we have not made the usual separation of our solution into a `+' and
a `-' branch, it is not immediately clear why these solutions are not ruled
out by the `no-go' theorems involving a dilaton potential \cite{nogo}; this
is presumably because those theorems exclude fluids with the equation of
state we have assumed. We also have not examined the relation of these
string-frame solutions to the corresponding Einstein-frame versions. These
issues await investigation. 

We thank M Gasperini, A Coley, R Tavakol, and the referee
for helpful comments, and
particularly J Lidsey for helpful discussions. DR wishes to thank Elaine
Kuok for her patience in checking many of the calculations in this paper. We
thank the NRF (South Africa) and Queen Mary College, London, for financial
support.

\appendix

\section*{Appendix A: Pre-Big Bang Evolution for Radiation}

For given $\rho _0$, it is convenient to define $y=\frac 23e^\phi \rho _0\ $
and change variables to $(t,y,\chi ).$ The equations (\ref{ev}) for $t>0$
become

\begin{equation}
\dot{y}=\chi y,\;\dot{\chi}=\frac{y-1}{(t+1)^2}+\frac \chi {2(t+1)},
\label{evplus}
\end{equation}
In the 2-dimensional sub-spaces $t=const$ with coordinates ($y,\chi ),$ the
curve $\gamma (t)$ has coordinates $(1,0)$ for all $t,$ and represents a set
of saddle points parametrised by $t$. To get exactly the desired dilaton
dynamics in the future ($\chi >0$, $e^\phi \rightarrow $ constant $
\Rightarrow \chi \rightarrow 0$ as $t\rightarrow \infty $), one must
restrict the initial conditions ($y_0$,$\chi _0$) to start precisely on the 
{\bf \ }stable branch of these saddle points, which intersects the surface $
t=0$ in a curve $(0,\gamma _{+}(\chi ),\chi )$ passing through the
exceptional point $\gamma _0=(0,1,0)$. One can obtain approximate solutions
by rewriting the second of eqs.(\ref{evplus}) in the form $\;$ 
\[
\left( \frac \chi {(1+t)^{1/2}}\right) ^{.}=\frac{y-1}{(t+1)^{5/2}} 
\]
Suppose $y$ is almost constant for $t>T_{+},$ implying $\chi $ is close to
zero then. Then we can integrate to get

\[
t>T_+ \Rightarrow \chi =-\frac 23\frac{y-1}{1+t}+C_{+}\sqrt{1+t} 
\]
where $C_{+}$ determines the magnitude of $\chi $ at time $T_{+.}$ The first
part decays away as desired, but the second part grows with time unless $
C_{+}=0;$ this is the fine-tuning required to attain the desired behaviour
of $\chi .$

To investigate $t<0$, it is again convenient to define $y=\frac 23e^\phi
\rho _0$ and change variables to $(t,y,\chi ).$ The equations for $t<0$
become

\begin{equation}
\dot{y}=\chi y,\;\dot{\chi}=y(-t+1)^2+\frac \chi {2(-t+1)}+\frac 1{(-t+1)^2},
\label{evminus}
\end{equation}
implying that $\dot{\chi}>0$ for all $t<0$; hence $\chi $ necessarily
decreases at all times in the past. The problem is that it can become
negative, because $\chi =0$ is not an invariant set of the equation. We want
a solution where $\chi $ remains positive for all time so that $\phi $
decreases for all time; this means we need $\chi $ to go to a positive value
or zero, but not to become negative, and $y$ to go to zero. . As in the
previous case one can obtain approximate solutions by rewriting the second
of eqs.(\ref{evminus}) in the form $\;$ 
\[
\left( \chi (1-t)^{1/2}\right) ^{.}=\frac 1{(-t+1)^{3/2}}(1+y(-t+1)^4). 
\]
Suppose 
\begin{equation}
y(-t+1)^4\ll 1~~{\rm for}~~t<T_{-}.  \label{inequ}
\end{equation}
Then we can ignore the second term on the right and integrate to get

\[
t<T_{-}\Rightarrow \chi =\frac 2{1-t}+\frac{C_{-}}{\sqrt{1-t}},~~y=y_0\frac
1{(1-t)^2}\exp (-2C_{-}\sqrt{1-t}) 
\]
where $C_{-},\;y_0$ represent the magnitude of $\chi ,\;y$ at time $T_{-\ }.$
This decays away as desired, and consistently preserves the inequality (\ref
{inequ}) for all earlier times because the exponential always dominates the
power law terms. The question then is whether for suitable initial
conditions we can attain this inequality at some time $T_{-},$ requiring $
\;y(T_{-})\ll (1-T_{-})^{-4}.$ We can satisfy this with $T_{-}=0$ if $
\;y_0=\frac 23e^{\phi_0}\rho _0\ll 1,$ i.e. $\phi_0\ll \ln (\frac 3{2\rho
_0}).$

\section*{Appendix B: Density evolution with exotic equation of state}

The `pre-big bang' evolution (\ref{t},\ref{t-}) 
implies $H$ and $\dot{H}$ in terms of $a$:

\begin{eqnarray}
\;t \geq 0:&H(a)=\frac 1{2a^2},\;\dot{H}(a)=\frac{-1}{2a^4},\;  \label{aa} \\
t \leq 0:&H(a)=\frac{a^2}2,\;\dot{H}(a)=\frac{a^4}2.
\end{eqnarray}
Assuming the exotic equation of state (\ref{exotic}) implied by setting $
\beta =0$ at all times, from (\ref{soln}) we find $\varphi $ in terms of $a$:
\begin{eqnarray}
t \geq 0:&\exp (\phi (a))=\exp (\phi _0)\frac a{a(1-2\chi _0)+2\chi _0},\;
\label{soln1} \\
t \leq 0:&\exp (\phi (a))=\exp (\phi _0)\frac{a^{5/2}}{a^{5/2}(1-\frac
25\chi _0)+\frac 25\chi _0\ },
\end{eqnarray}
and from (\ref{chi1},\ref{neg}) we find $\chi $ in terms of $a$:
\begin{eqnarray}
t \geq 0:&\chi (a)=\frac{\chi _0}{a^2\left( 2\chi _0+(1-2\chi _0)\ a\right) 
},  \label{chi11} \\
t \leq 0:&\;\chi (a)=-\ \frac{5a\chi _{_0}}{2\chi _{_0}+(5-2\chi
_{_0})a^{5/2}}.
\end{eqnarray}
A particularly simple case occurs when $\chi _0=\frac 14$. Then 
\begin{eqnarray}
t \geq 0: &\exp (\phi (a))=\exp (\phi _0)\frac{2a}{a+1},\;\\
t \leq 0:&\exp (\phi
(a))=\exp (\phi _0)\frac{10a^{5/2}}{9a^{5/2}+1\ }.  \label{phi2a}
\end{eqnarray}
and
\begin{eqnarray}
t \geq 0:&\chi (a)=\frac 1{2a^2\left( 1\ +\ a\right) },\\
\;t \leq 0:&\chi (a)=-\ 
\frac{5a\ }{2(1\ +9a^{5/2})}.  \label{chi2a}
\end{eqnarray}

Now $\rho (t)$ is determined by (\ref{ccons}); using the above expressions,
for $t>0$ and  $\chi _0=\frac 14$ this becomes
\[
\frac{d\rho }{da}=-\frac 3{a^5}-\frac 3{4a^6(1+a)}
\]
which can be solved to give
\[
\rho (a)=C+\frac 3{20a^5}+\frac 9{16a^4}+\frac 1{4a^3}-\frac 3{8a^2}+\frac
3{4a}+\frac 34\ln (\frac a{1+a}).
\]
This implies $\rho (t)\rightarrow C+\frac{107}{80}-\frac 34\ln 2=C+0.81764...
$ as $t\rightarrow 0_{+}$ and $\rho (t)\rightarrow C$ as $t\rightarrow
\infty ;$ hence choosing $C=0,$ $\rho (t)\rightarrow 0.81764...$ as $
t\rightarrow 0_{+}$ and $\rho (t)\rightarrow 0$ as $t\rightarrow \infty .$
Also $p/\rho \rightarrow 1/3$ as $t\rightarrow \infty .$ The expression for $
V(\varphi )$ in this case follows on putting this into(\ref{V}) and using (
\ref{solnst}), (\ref{t}), and the various expressions above. Similar (more
complicated) expressions can be obtained for $t<0.$


\begin{thebibliography}{99}

\bibitem{Gasperini}  Gasperini, M. Lectures at ``VI Seminario Nazionale di
Fisica Teorica'' (1999), see hep-th/9907067. See also the references at\\ 
http://www.to.infn.it/\symbol{126}gasperin/

\bibitem{Lidsey}  Lidsey, J E, Wands, D, and Copeland, E J. `Superstring
Cosmology' (1999). See hep-th/9909061.

\bibitem{Veneziano}  Gasperini, M. and G. Veneziano. 1993 Astropart. Phys. 1
317.

\bibitem{Lidsey1}  Lidsey J E, Liddle A R, Kolb E W, Copeland E J, Barreiro
T, and Abney M. `Reconstructing the Inflaton Potential: An Overview'. Rev
Mod Phys 69: 373 (1977).

\bibitem{Saini}  Saini T D, Raychaudhury S, Sahni V, and Starobinsky A A.
`Reconstructing the cosmic equation of state from Supernovae distances'
(1999). See astro-ph/9910231.

\bibitem{Ellis}  Ellis, G F R, and Madsen, M S. 1991. Class. Quantum Grav.
8, 667-676.

\bibitem{wain}  Wainright J and Ellis G\ F\ R. 1997. {\it Dynamical Systems
in Cosmology}. Cambridge University Press.

\bibitem{nogo}  Kaloper, N, Madden, R, and Olive, K A. 1995. Towards a
Singularity Free Inflationary Universe? hep-th/9506027; 
Axions and the Graceful Exit Problem in String Cosmology. 
hep-th/9510117.

\bibitem{nogo1}  Brustein, R, and Madden, R. 1997. Graceful exit and energy
conditions in string cosmology. {\it Phys.Lett}. {\bf B410} (1997) 
110-118. hep-th/9702043.

\bibitem{trans1}
Brustein, R., and Madden, R.
A Model of Graceful Exit in String Cosmology
{\it Phys.Rev}. {\bf D57} (1998) 712-724.
hep-th/9708046.

\bibitem{trans2} 
Foffa, S., Maggiore, M., and Sturani, R.
Loop corrections and graceful exit in string cosmology
{\it Nucl.Phys.} {\bf B552} (1999) 395-419.
hep-th/9903008.

\bibitem{trans3} 
Cartier, C., Copeland, E.J. and Madden, R.
The graceful exit in pre-big bang string cosmology
{\it JHEP} {\bf 0001} (2000), 035.
hep-th/9910169.


\end{thebibliography}
\end{document}